\documentclass[pdflatex,sn-mathphys-num]{sn-jnl}
\usepackage{graphicx}%
\usepackage{multirow}%
\usepackage{amsmath,amssymb,amsfonts}%
\usepackage{amsthm}%
\usepackage{mathrsfs}%
\usepackage[title]{appendix}%
\usepackage{xcolor}%
\usepackage{textcomp}%
\usepackage{manyfoot}%
\usepackage{booktabs}%
\usepackage{algorithm}%
\usepackage{algorithmicx}%
\usepackage{algpseudocode}%
\usepackage{listings}%
\begin{document}

\title{\textbf{Nonlocal-nonlinear phonon polaritons}}


\author*[1]{\fnm{Gonzalo} \sur{\'Alvarez-P\'erez}}\email{gonzalo.alvarezperez@iit.it}
\author[2]{\fnm{Line} \sur{Jelver}}
\author[3]{\fnm{Alexander} \sur{Paarmann}}
\author[4]{\fnm{Simone} \sur{De Liberato}}

\affil[1]{Istituto Italiano di Tecnologia, Center for Biomolecular Nanotechnologies, Via Barsanti 14, 73010 Arnesano, Italy}

\affil[2]{POLIMA—Center for Polariton-driven Light–Matter Interactions, University of Southern Denmark, Campusvej 55, DK-5230 Odense M, Denmark}

\affil[3]{Department of Physical Chemistry, Fritz Haber Institute of the Max Planck Society, Faradayweg 4-6, 14195, Berlin, Germany}

\affil[4]{Istituto di Fotonica e Nanotecnologie, Consiglio Nazionale delle Ricerche (CNR), Piazza Leonardo da Vinci 32, Milano, 20133, Italy}


\abstract{When light is confined to deeply subwavelength volumes, optical responses can become nonlocal. Yet how nonlocality reshapes nonlinear light–matter interactions remains largely unexplored, particularly in dielectrics. Here, we theoretically uncover a new class of nonlinear optical processes arising from the interplay between optical-phonon propagation and ionic-bond anharmonicity in polar crystals. We show that, unlike conventional local nonlinearities, this mechanism enables second-order responses even in centrosymmetric materials. Moreover, these responses can be tuned by the confinement and resonantly enhanced near optical-phonon resonances. Using density functional perturbation theory, we quantify the resulting nonlocal second-order susceptibilities in representative polar crystals. Our results establish nonlocal-nonlinear optics as a new regime of light–matter interaction in dielectrics.}


\maketitle


In the mid-infrared, phonon polaritons (PhPs)---quantum superpositions of photons and lattice vibrations in polar crystals---can confine light to deeply subwavelength volumes through coherent ionic oscillations with low optical loss \cite{Caldwell15low}. In nanoresonators, nanoantennas, or two-dimensional layers, PhPs can achieve extremely small mode volumes and strong near-field enhancement \cite{HerzigSheinfux2024_high,Klein25_nanometer}. Extreme confinement can also be achieved in anisotropic polar crystals, including hyperbolic van der Waals (vdW) and bulk materials, where the directional propagation of PhPs further compresses electromagnetic fields \cite{Zhou2026_Fundamental,kowalski25_ultraconfined}. At such extreme levels of confinement, the lattice displacement is no longer uniform across neighboring unit cells, giving rise to spatial gradients in the displacement field. As a result, PhPs confined to only a few interatomic spacings acquire an explicit wavevector dependence, or nonlocality \cite{Monticone_25_nonlocality}. First studied in plasmonics \cite{Mortensen_21_mesoscopic,ciraci2012probing,Rajabali2021_polaritonic}, optical nonlocality extends over tens of nanometers and gives rise to even richer phenomena in polar dielectrics. These include reduced confinement, Kreibig-like broadening, quantized longitudinal optical (LO) phonon resonances, nonlocal screening, and shifted epsilon-near-zero resonances \cite{Gubbin21_impact,Heiden_25_bypassing,Gubbin20_optical,ma_26_probing,gubbin22_Quantum}. Unlike the blueshift of nonlocal plasmons, the redshift of optical phonons brings nonlocal corrections into resonance with localized modes. This produces hybrid longitudinal-transverse PhPs that have been both predicted theoretically \cite{Gubbin20_optical,gubbin22_Quantum,Gubbin2023_electrical} and observed experimentally \cite{Gubbin2019_Hybrid,Ratchford2019_Controlling}.

The same confinement that drives nonlocal linear response also makes PhPs an ideal platform for nonlinear optics, where strong field enhancement amplifies processes such as second-harmonic generation (SHG) \cite{Paarmann16effects} and four-wave mixing \cite{Gubbin_18_theory}. Polar dielectrics are particularly attractive because their nonlinear response originates from lattice anharmonicity in the terahertz and mid-infrared, while the same lattice supports resonant PhPs. Consequently, resonantly enhanced second-order processes have been demonstrated in SiC nanopillars \cite{Razdolski16_resonant,Razdolski18_second}, critically coupled PhPs \cite{Passler17_second}, and grating-coupled hybrid plasmon--phonon polaritons \cite{Kohlmann22_second}, with zone-folded phonons in 4H- and 6H-SiC providing particularly strong enhancement of $\chi^{(2)}$ \cite{Paarmann16effects,Razdolski16_resonant}. Beyond SHG, infrared-visible sum-frequency generation (SFG) microscopy has enabled the observation of phonon-enhanced $\chi^{(2)}$ \cite{Mueller26_full} and the direct imaging of localized \cite{niemann2022long} and propagating \cite{niemann2024spectroscopic} PhPs with subdiffraction-limited spatial resolution. Together, these studies establish PhPs as a powerful platform for enhanced second-order nonlinear optics. Yet they—and generally conventional nonlinear optics in polar dielectrics—treat the nonlinear susceptibility as spatially local \cite{boyd2003nonlinear}, overlooking the interplay between nonlocality and nonlinearity despite their common origin in strong light-matter coupling.

\begin{figure}[!ht]
\centering 
\includegraphics[width=0.7\textwidth]{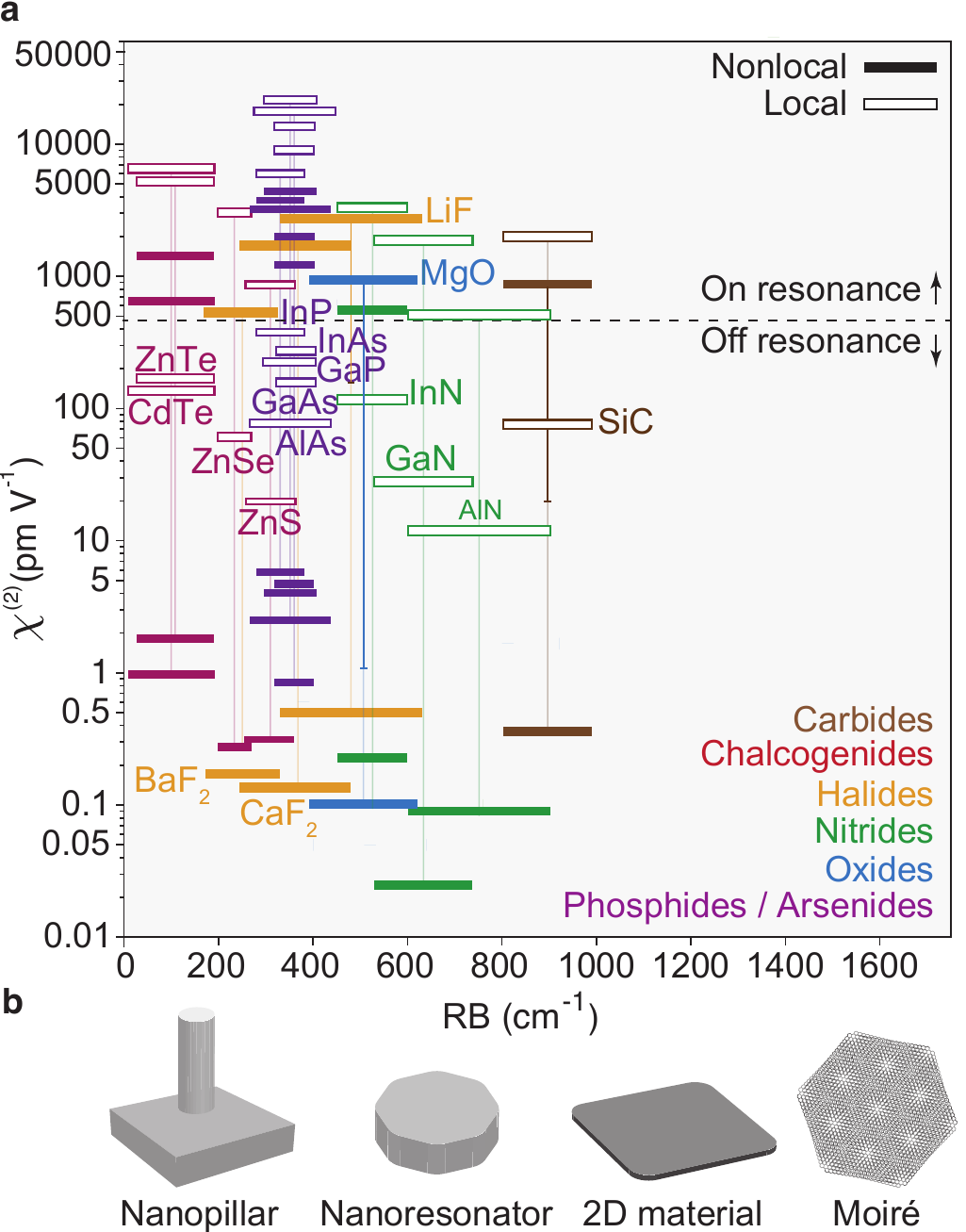}
\caption{\label{fig1}
\textbf{Second-order nonlinear susceptibilities $\chi^{(2)}$ and RBs of several polar dielectrics.}
\textbf{(a)} Local and nonlocal $\chi^{(2)}$ as a function of the RB frequency for representative polar crystals, both off and on resonance (at $k = 150k_0$) Vertical guide-to-the-eye lines link values for each material. Error bars for LiF, MgO, and SiC indicate the spread of DFPT values across polarization channels and propagation directions, rather than numerical uncertainty, as will be discussed below.
\textbf{(b)} Schematic of polar dielectric nanostructures supporting PhPs.
}
\end{figure}

Interestingly, within this local framework, $\chi^{(2)}$ vanishes in centrosymmetric media because inversion symmetry forces the induced polarization from a uniform field to cancel. The established exception is the electric-quadrupole (more generally multipolar) response, where field gradients generate a weak second-order polarization proportional to the gradient rather than the local field \cite{Bloembergen68_Optical,Tang26_forbidden}. Physically, spatially varying fields displace ions within a unit cell unequally, lifting the cancellation imposed by inversion symmetry. This mechanism underlies bulk quadrupolar contributions and surface nonlinearities, but in conventional optics it remains weak because electromagnetic fields vary over length scales far larger than the unit cell. Recent work in plasmonic metals and heavily doped semiconductors has shown that material nonlocality can strongly modify and enhance nonlinear optical processes \cite{de2021free,hu2024low,Rossetti2025_control,Hu25_Modulating,AlvarezPerez2025_ultrahigh}. Unlike conventional quadrupolar nonlinearities, these effects arise from the spatially dispersive response of the material itself rather than from electromagnetic field gradients. Motivated by the strong influence of nonlocality on linear PhP physics, we ask whether lattice nonlocality can similarly create new nonlinear optical channels. In this work, we show that it does. We identify a second-order nonlinear mechanism arising from the interplay between lattice nonlocality and anharmonicity. Under strong confinement, spatial variations of the nonlocal lattice response directly generate a second-order polarization absent in local theory. Unlike conventional quadrupolar processes, which are driven by field gradients, this effect is driven by gradients of the lattice polarization itself and therefore represents a fundamentally distinct bulk nonlinear mechanism. We develop a general theory of nonlocal lattice nonlinearity and show that, in centrosymmetric polar dielectrics where dipolar $\chi^{(2)}$ is symmetry-forbidden, it becomes the dominant bulk source of SHG, enhanced by both strongly confined PhPs and resonances near optical phonons. These results establish nonlocal-nonlinear PhPs as a new route to second-order nonlinear optics under extreme confinement.


Figure~\ref{fig1}a surveys value of the second-order susceptibility, $\chi^{(2)}$, across the Reststrahlen bands (RBs)—the spectral region between $\omega_T$ and $\omega_L$, the transverse- and longitudinal-optical phonon bands in which surface PhPs are supported—for a broad range of polar crystals. The materials span chalcogenides, III--V semiconductors, nitrides, oxides, carbides, and halides, illustrating the diversity of phonon frequencies and infrared nonlinear responses available across polar dielectrics. Non-centrosymmetric compounds exhibit bulk dipolar susceptibilities reaching several hundred pm/V throughout the mid- and far-infrared \cite{Shoji97_Absolute,Skauli03_Improved,boyd2003nonlinear,sutherland2003handbook,Sanford05_measurement}, whereas centrosymmetric materials such as MgO, LiF, CaF$_2$, and BaF$_2$ have no local bulk $\chi^{(2)}$ by symmetry. These values provide a benchmark for the nonlocal nonlinear response that we introduce in this work. We will show that nonlocality yields a resonant correction to the conventional susceptibility in non-centrosymmetric crystals, but becomes the dominant bulk source of SHG in centrosymmetric materials. First-principles calculations and analytical estimates predict resonantly enhanced nonlocal susceptibilities comparable to the largest phonon-mediated infrared nonlinearities reported to date, with fluorides—particularly LiF, followed by CaF$_2$ and BaF$_2$—emerging as especially promising owing to their proximity to the doubly resonant condition $\omega_L\approx2\omega_T$, as we will see below. While similarly large nonlinearities have previously been realized mainly in layered vdW crystals and ferroelectric or soft-mode oxides through anisotropic, multimode, or strongly anharmonic lattice dynamics \cite{Mueller26_full,Bergeron2023_Probing},
the giant nonlinearities predicted here instead arise from the intrinsically nonlocal lattice response coupled to phonon anharmonicity under extreme spatial confinement.

\section*{Theory of nonlocal SHG in polar crystals}

In our formalism, we treat the crystal as a continuous medium with ionic mass and charge densities $\rho$ and $\mu$, and a continuous relative displacement field $X_\alpha(\mathbf{r},t)$ between the positive and negative sublattices \cite{Born54_Dynamical}. Its dynamics follow
\begin{equation}
    \rho\ddot{X}_\alpha+\rho\gamma_0\dot{X}_\alpha=-\frac{\delta U[\{X_\alpha\}]}{\delta X_\alpha}+\mu E_\alpha,
    \label{euler_lagrange}
\end{equation}
where $\gamma_0$ is a damping rate and $U[\{X_\alpha\}]=\sum_i U_i[\{X_\alpha\}]$ is a mechanical potential, playing the role of electron pressure in the analogous hydrodynamic description of metals \cite{Mortensen_21_mesoscopic,ciraci2012probing}. We take $U$ to depend only on the lattice displacement, so that all nonlocality and anharmonicity are of lattice origin (see Methods for the general field-gradient formulation and the linear nonlocal limit $U\approx U_2$, which recovers the framework of \cite{Gubbin20_optical}). Higher-order terms $U_3, U_4, \ldots$ capture the anharmonicities responsible for nonlinear responses. In particular, the cubic term,
\begin{equation}
U_3 = \frac{\rho}{6}\int\!\frac{d^3k\,d^3k'}{(2\pi)^6}\,
g_{\alpha\beta\gamma}(\mathbf{k},\mathbf{k}')\,
\tilde{X}_\alpha(-\mathbf{k}-\mathbf{k}')\,
\tilde{X}_\beta(\mathbf{k})\,
\tilde{X}_\gamma(\mathbf{k}'),
\label{eq:U3}
\end{equation}
accounts for second-order processes. The cubic force-constant vertex $g_{\alpha\beta\gamma}(\mathbf{k},\mathbf{k}')$---the Fourier-space third derivative of the ionic potential---physically encodes the leading-order asymmetry of the restoring force felt by a displaced ion's neighbors. It also governs phonon-phonon scattering, finite phonon lifetimes, and thermal transport in the phonon eigenmode basis. We treat $\gamma_0$ as phenomenological, neglecting any double counting with the linewidth contribution from $U_3$.

Substituting $U_3$ into Eq.~\eqref{euler_lagrange} via $-\delta U_3/\delta X_\alpha$ yields a nonlinear source term, coupling different wavevectors and polarizations. Higher-order terms $U_4, U_5, U_6, \ldots$ give rise to further nonlinear effects such as third-harmonic generation, optical Kerr effects, and cascaded processes. These nonlinear processes have been studied extensively in the nonlocal hydrodynamic model of metals, where spatial-gradient terms underpin strong and tunable nonlinear optical responses \cite{de2021free, Rossetti2025_control, hu2024low, Hu25_Modulating, AlvarezPerez2025_ultrahigh, Hallman2025_HHG}. We expect analogous terms to be accessible in polar crystals, though their analysis lies beyond the scope of this work.

\begin{figure}[!ht]
\centering
\includegraphics[width=0.7\textwidth]{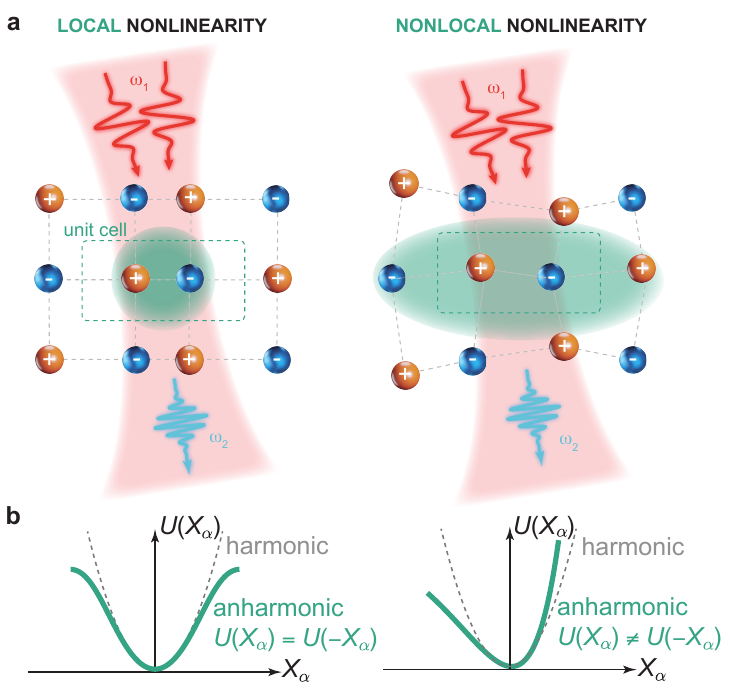}
\caption{\label{fig2}
\textbf{Local versus nonlocal origins of second-order optical nonlinearity in a crystal lattice.}
\textbf{(a)} In the local picture \textbf{(left)}, the lattice response is identical from one unit cell to the next, so the second-order polarization at frequency $\omega_2$ is determined solely by the local response to the incident field at $\omega_1$. In the nonlocal picture \textbf{(right)}, the lattice response varies across neighboring unit cells, giving rise to a spatially-dispersive (nonlocal) second-order polarization.
\textbf{(b)} Potential energy landscape illustrating inversion symmetry: the local cubic anharmonicity $U^{(0)}_3$ satisfies $U(X_\alpha) = U(-X_\alpha)$ and is therefore forbidden in centrosymmetric crystals \textbf{(left)}, while the gradient-coupled term $U^{(1)}_3$ satisfies $U(X_\alpha) \neq U(-X_\alpha)$ under inversion and is allowed \textbf{(right)}.
}\end{figure}

Near the $\Gamma$ point, $g_{\alpha\beta\gamma}[(\mathbf{k},\mathbf{k}')=\textbf{0}]$ admits the Taylor expansion
\begin{align}
g_{\alpha\beta\gamma}[(\mathbf{k},\mathbf{k}')=\textbf{0}]
&\approx g^{(0)}_{\alpha\beta\gamma}
+ A_{\alpha\beta\gamma\mu}\,k_\mu
+ B_{\alpha\beta\gamma\mu}\,k'_\mu \nonumber \\
&+ \tfrac{1} 
{2}\bigl[C_{\alpha\beta\gamma\mu\nu}\,k_\mu k_\nu
+ D_{\alpha\beta\gamma\mu\nu}\,k_\mu k'_\nu
+ E_{\alpha\beta\gamma\mu\nu}\,k'_\mu k'_\nu\bigr]
+ \mathcal{O}(k^3),
\label{eq:g_expansion}
\end{align}
where the coefficients are the successive derivatives of $g$ evaluated at $(\mathbf{k},\mathbf{k}')=\mathbf{0}$: $A_{\alpha\beta\gamma\mu}=\partial g/\partial k_\mu|_0$, $B_{\alpha\beta\gamma\mu}=\partial g/\partial k'_\mu|_0$, and $C$, $D$, $E$ the corresponding second-derivative blocks, with the factor $1/2$ absorbing the mixed-derivative combinatorics. Here and throughout, $\mu$ denotes a Cartesian index (as $\alpha$, $\beta$, $\delta$, $\nu$ \ldots), not to be confused with the effective charge density $\mu$ in Eq.~\eqref{euler_lagrange}. Truncating at zeroth order $g_{\alpha\beta\gamma}\approx g^{(0)}_{\alpha\beta\gamma}$, which reduces to the well-known second-order susceptibility $\chi^{(2)}$ and thus recovers the limit in which the nonlinear response is purely local (Fig. \ref{fig2}a, left). Retaining the $\mathcal{O}(k)$ and $\mathcal{O}(k^2)$ terms introduces nonlocal corrections to the anharmonic force that grow with the polariton wavevector $k$ and constitute the main focus of this work (Fig.~\ref{fig2}a, right).

\section*{Selection rules for local and nonlocal PhP nonlinearities}

\begin{figure*}[!ht]
\includegraphics[width=\textwidth]{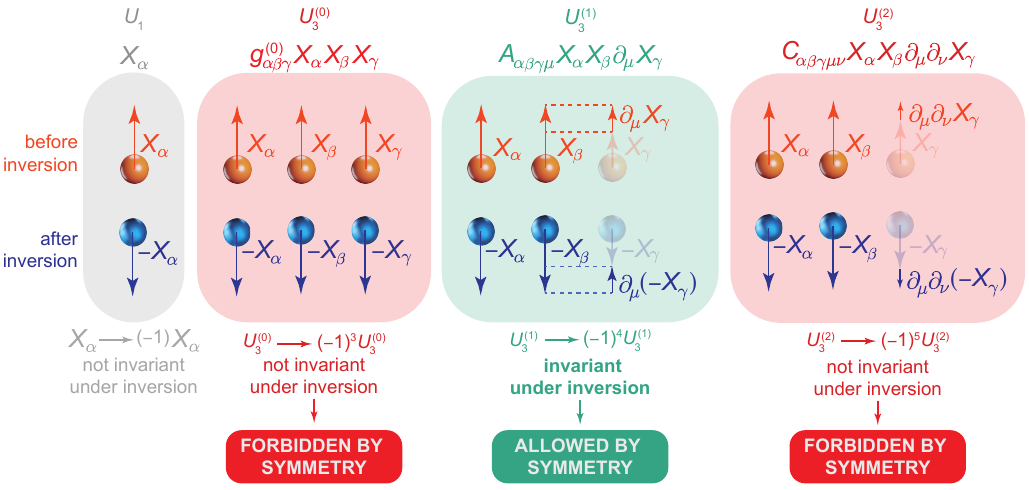}
\caption{\label{fig3}
\textbf{Schematic representation of the symmetry selection rules for the three leading anharmonic energy contributions of nonlocal-nonlinear PhPs under spatial inversion.} $U^{(0)}_3$ (zero spatial derivatives) and $U^{(2)}_3$ (two spatial derivatives) acquire a sign change and are forbidden; $U^{(1)}_3$ (one spatial derivative) is invariant and is therefore the leading-order nonlinear contribution in centrosymmetric materials.
}\end{figure*}

Not all terms are allowed in Eq. \eqref{eq:g_expansion} for all materials. We therefore now investigate how symmetry constrains the anharmonic energy functional $U_3$. Transforming to real space via $k_\mu\tilde{X}_\beta\leftrightarrow-i\partial_\mu X_\beta$, $k_\mu k_\nu\tilde{X}_\beta\leftrightarrow-\partial_\mu\partial_\nu X_\beta$, and using the permutation symmetry $g_{\alpha\beta\gamma}(\mathbf{k}, \mathbf{k}')=g_{\alpha\gamma\beta}(\mathbf{k}',\mathbf{k})$ (which implies $B_{\alpha\beta\gamma\mu}=A_{\alpha\gamma\beta\mu}$), the cubic functional decomposes as $U_3=\sum_{i=0}^{2}U_3^{(i)}$, with $U_3^{(0)} = \frac{\rho}{6}\int d^3r \,g^{(0)}_{\alpha\beta\gamma} X_\alpha X_\beta X_\gamma$, 
$U_3^{(1)} = \frac{\rho}{3}\int d^3r \, A_{\alpha\beta\gamma\mu}X_\alpha X_\beta \partial_\mu X_\gamma$, 
$U_3^{(2)} = -\frac{\rho}{12}\int d^3r \, \Big[C_{\alpha\beta\gamma\mu\nu} X_\alpha X_\gamma \partial_\mu\partial_\nu X_\beta + D_{\alpha\beta\gamma\mu\nu} X_\alpha \partial_\mu X_\beta \partial_\nu X_\gamma + E_{\alpha\beta\gamma\mu\nu}X_\alpha X_\beta \partial_\mu\partial_\nu X_\gamma \Big]$. These three terms form a systematic gradient expansion: $U_3^{(0)}$ is the local cubic anharmonicity responsible for conventional bulk $\chi^{(2)}$; $U_3^{(1)}$ is the leading nonlocal correction coupling the displacement field to its first spatial gradient; and $U_3^{(2)}$ contains second-order gradient corrections (Fig. \ref{fig2}b).

The corresponding nonlinear force densities $\xi^{(n)}_\alpha=-\delta U_3^{(n)}/\delta X_\alpha$ follow by functional differentiation. From $U_3^{(0)}$: $\xi^{(0)}_\alpha = -\frac{1}{2}\rho\, g^{(0)}_{\alpha\beta\gamma} X_\beta X_\gamma$,
generating the familiar local $\chi^{(2)}$ response. From $U_3^{(1)}$, integrating by parts:
$\xi^{(1)}_\alpha = -\frac{\rho}{3} \Big[ A_{\alpha\beta\gamma\mu}X_\beta\partial_\mu X_\gamma + A_{\beta\alpha\gamma\mu}X_\beta\partial_\mu X_\gamma - A_{\beta\gamma\alpha\mu}\partial_\mu(X_\beta X_\gamma) \Big]$. For a plane-wave fundamental with wavevector $\mathbf{k}$, the three force densities scale as $\xi^{(0)} \sim g^{(0)}X^2$, $\xi^{(1)} \sim AkX^2$, $\xi^{(2)} \sim Ck^2X^2$, so the corresponding nonlinear susceptibilities scale as $\chi^{(2)} \propto 1,\, k,\, k^2$ at successive orders.

In centrosymmetric crystals, under inversion symmetry, $X_\alpha(\mathbf{r})\to -X_\alpha(-\mathbf{r})$, so a cubic monomial containing $n_\partial$ spatial derivatives transforms as $(-1)^{3+n_\partial}$; invariance requires $n_\partial$ odd. Equivalently, in momentum space, invariance requires $g_{\alpha\beta\gamma}(\mathbf{k},\mathbf{k}') = g_{\alpha\beta\gamma}(-\mathbf{k},-\mathbf{k}')$, so the cubic vertex is an odd function of $(\mathbf{k},\mathbf{k}')$. This immediately forbids $U_3^{(0)}$ ($n_\partial=0$) and $U_3^{(2)}$ ($n_\partial=2$), forcing $g^{(0)}$, $C$, $D$, and $E$ to vanish, while $U_3^{(1)}$ ($n_\partial=1$) is allowed, as schematically illustrated in Fig. \ref{fig3}. Consequently, $U_3 = U_3^{(1)}$ to leading order in the long-wavelength expansion. 

The vanishing of $g^{(0)}$ simply reflects the familiar argument that a perfectly symmetric potential has no cubic term: a spatially uniform field displaces all ions identically and produces no net second-order polarization---the standard reason why bulk $\chi^{(2)}=0$ in centrosymmetric media. In contrast, the $U_3^{(1)}$ term couples the displacement of an ion not to its own amplitude, but to the amplitude difference between neighboring ions, i.e., to the local strain or field gradient (Fig. \ref{fig2}b). When a deeply confined PhP drives the crystal, neighboring unit cells experience slightly different field amplitudes because the field varies appreciably over a few lattice spacings. This spatial asymmetry effectively tilts the potential (Fig. \ref{fig2}b, right), generating a net second-harmonic polarization even though the crystal structure is centrosymmetric. Representative materials where these nonlinearities can be probed are collected in Table~1 of the Supplementary Information.

The mechanism is therefore an intrinsic bulk nonlocal response---it vanishes in the local limit, grows linearly with $k$, and therefore can be significantly enhanced under the extreme sub-diffractional confinement characteristic of PhPs. As such, it should be distinguished from surface SHG \cite{Jha65_Nonlinear,Brown65_nonlinear}, which originates from the explicit breaking of inversion symmetry at a crystal termination and remains finite as $k\to 0$. The two contributions can coexist at an interface supporting PhPs---for example, at the surface of a MgO or LiF nanostructure---but scale differently with confinement: the surface term is $k$-independent and localized to the interface, while the nonlocal bulk term grows as $k$ and is distributed throughout the mode volume, and is therefore expected to dominate under strong PhP confinement. 

Moreover, in the local limit, where the displacement field is proportional to the electric field ($X_\alpha \propto E_\alpha$), these terms reduce to the phenomenological quadrupolar nonlinearities reported in early studies of nonlinear optics in centrosymmetric media \cite{Bloembergen68_Optical}, which give rise to nonlinear polarizations of the form $P^{(2)} \propto E\nabla E$. Here, however, the nonlinearity originates microscopically from the gradient-dependent anharmonic term $U^{(1)}_3$, which is allowed by the spatial gradients of neighboring displacements breaking inversion symmetry. Under strong PhP confinement, finite phonon propagation and the participation of longitudinal phonon modes cause $X_\alpha$ and $E_\alpha$ to develop distinct spatial profiles, making the lattice-dynamical origin of the nonlinearity explicit. The resulting response therefore retains the symmetry and $k$-linear scaling of conventional quadrupolar nonlinearities, but its magnitude is governed by phonon resonances and by anharmonic force constants that can be computed directly from the third-order dynamical matrix. 

\section*{First-principles estimates of the local and nonlocal $\chi^{(2)}$}

\begin{figure*}[!ht]
\centering
\includegraphics[width=\textwidth]{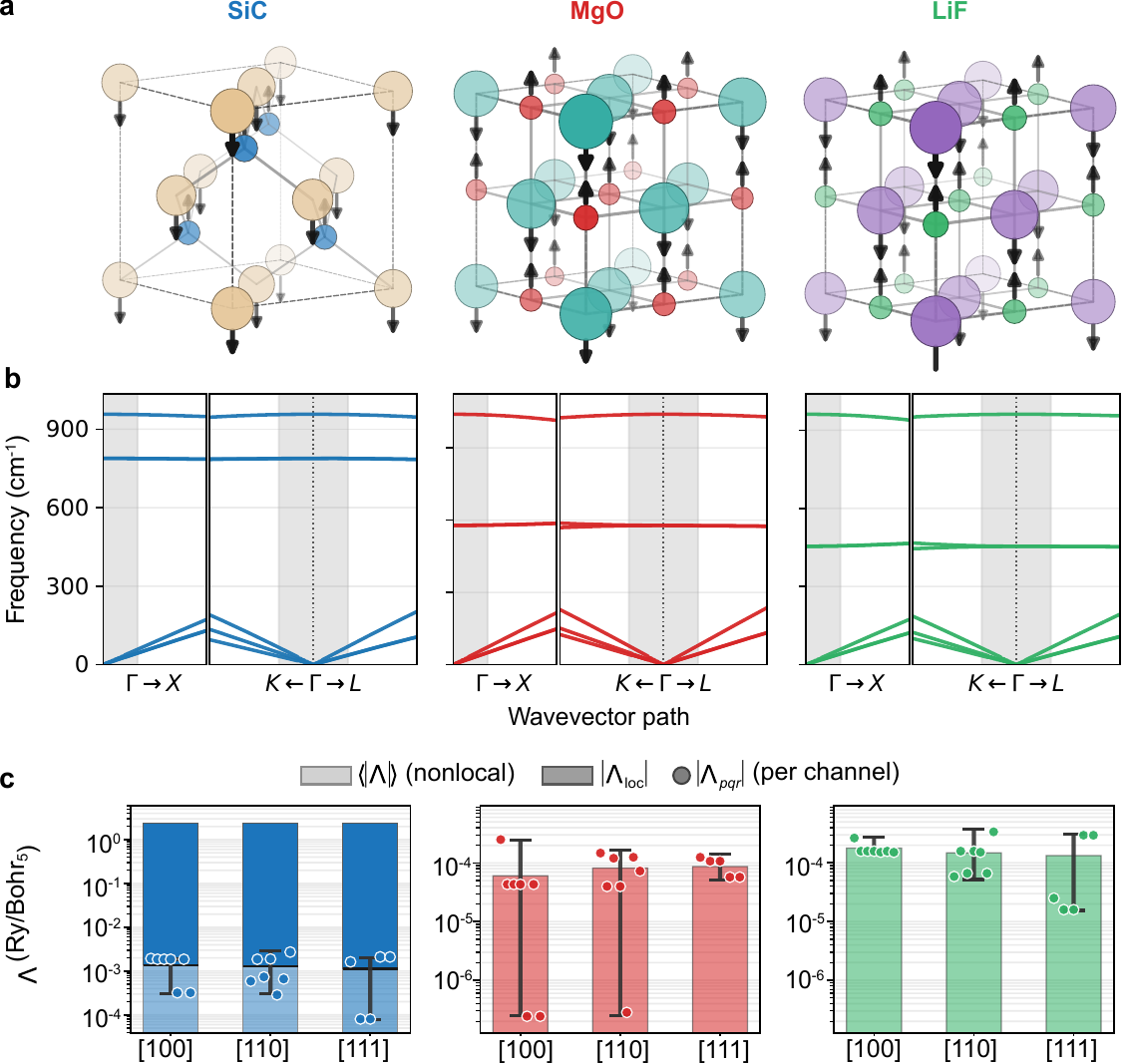}
\caption{\label{fig4}
\textbf{First-principles phonon dispersions and nonlocal anharmonic coefficients for three representative polar crystals.}
\textbf{(a)} Relaxed conventional unit cells of 3C-SiC (\textbf{left}), MgO (\textbf{middle}) and LiF (\textbf{right}). Arrows depict the $\Gamma$-point TO phonon displacement pattern at fixed amplitude for clarity. 
\textbf{(b)} Harmonic phonon band structures of cubic 3C-SiC, MgO, and LiF calculated using DFPT with the PBEsol functional, plotted along high-symmetry paths in the Brillouin zone. Grey shading marks the small-wavevector fitting window ($q^2 \leq 0.005\,(2\pi/a)^2$) used to extract the nonlocal harmonic coefficients $\beta_L$ and $\beta_T$. The full harmonic phonon band structures are shown in the Supplementary Information.
\textbf{(c)} Magnitude of the nonlocal anharmonic tensor components along the $[100]$, $[110]$, and $[111]$ propagation directions $\hat{q}$ of the PhP for each crystal. Bars show the channel-averaged nonlocal norm $\langle|\Lambda|\rangle$ (grey), while dots indicate the individual per-channel contributions $|\Lambda_{pqr}(\hat{q})|$, where $p,q,r \in \{L, T_1, T_2\}$ label phonon-polarization channels relative to $\hat{q}$. Note that for centrosymmetric MgO and LiF the local cubic term vanishes identically, leaving $U_3^{(1)}$ as the dominant source of second-order nonlinearity. The insets show a zoomed in version of the small-$k$ regions.
}
\end{figure*}

Now we calculate the nonlocal second-order susceptibility using first-principles density-functional perturbation theory (DFPT). The calculations, performed with Quantum ESPRESSO and the D3Q package \cite{Giannozzi_2009,Giannozzi_2017,Paulatto_2013}, provide both the harmonic phonon properties and the third-order dynamical-matrix elements required to extract the nonlocal susceptibility (see Methods). Figure~\ref{fig4}b shows the resulting phonon dispersions together with the $\Gamma$-point TO phonon mode of each crystal (the full spectra are shown in the Supplementary Information). Because the DFPT nonlocal response depends on both propagation direction and phonon polarization, we report the range of projected values obtained for different polarization channels along the $[100]$, $[110]$, and $[111]$ directions (Fig.~\ref{fig4}c). The spread thus reflects the intrinsic anisotropy of the projected nonlocal tensor across propagation directions and phonon-polarization channels, rather than numerical uncertainty. Here, these directions denote the propagation direction $\hat{q}$ of the small finite wavevector in the D3Q triplet $(q,0,-q)$, i.e. the PhP propagation direction, rather than the optical field-polarization directions conventionally used to label the crystal-frame tensor components $\chi^{(2)}_{pqr}$ (see Methods). We complement the DFPT calculations with dimensional estimates requiring only tabulated phonon frequencies and lattice parameters, enabling a broad survey of materials beyond those treated explicitly with DFPT, as shown in Fig. \ref{fig1}.

We first benchmark our framework against well-established local $\chi^{(2)}$ of non-centrosymmetric crystals. For 3C-SiC, the DFPT calculation using the continuum-limit anharmonic vertex $g^{(0)}_{\alpha\beta\gamma}$ extracted from the full third-order dynamical matrix yields $|\chi^{(2)}_0|^\mathrm{DFPT}=83.4\ \mathrm{pm/V}$, in good agreement with experiment \cite{Paarmann16effects,Gubbin17_theory}. At $k=150k_0$, the leading nonlocal contribution ranges from $9.5\times10^{-3}$ to $0.35\ \mathrm{pm/V}$, reflecting the strong channel dependence of $\Lambda_{pqr}(\hat{q})$ (see Methods). The crossover wavevector at which local and nonlocal contributions become comparable,
$k^*=8.4\times10^7~\mathrm{cm^{-1}}$, lies far beyond experimentally achievable PhP confinement.
Even so, symmetry can suppress the bulk local $\chi^{(2)}$ for selected polarization or azimuthal
configurations, allowing the nonlocal contribution to dominate. More importantly, the local
$\chi^{(2)}$ can itself be exploited as a heterodyne reference to amplify the nonlocal signal.

These DFPT results are reproduced by a simple dimensional analysis. The anharmonic force-constant vertex $g^{(0)}_{\alpha\beta\gamma}$ in Eq.~\eqref{eq:U3} has units $ML^{-4}T^{-2}$, and successive powers of wavevector add one power of length, giving $[A]=[B]=ML^{-3}T^{-2}$, $[C]=[D]=[E]=ML^{-2}T^{-2}$. Using the natural microscopic scales $\rho\omega_T^2$ and lattice spacing $a$ ($g^{(0)}\sim\rho\omega_T^2a^{-1}$, $A\sim B\sim\rho\omega_T^2$, $C\sim D\sim E\sim\rho\omega_T^2a$), the nonlinear sources scale as $\xi^{(0)}\sim g^{(0)}$, $\xi^{(1)}\sim g^{(0)}ka$, $\xi^{(2)}\sim g^{(0)}(ka)^2$, each successive order suppressed by $ka\approx3.5\times10^{-2}$ for $k=150k_0$ and $\lambda=8\,\mu$m, values routinely achieved in PhP nanoresonators \cite{HerzigSheinfux2024_high,Klein25_nanometer,kowalski25_ultraconfined}. The corresponding local susceptibility follows from
\begin{equation}
\left[\chi^{(2)}_{0}\right]_{\alpha\beta\gamma}(\omega_3;\omega_1,\omega_2) = -\frac{\mu^3 g^{(0)}_{\alpha\beta\gamma}}{2\varepsilon_0\rho^3\Delta(\omega_3)\Delta(\omega_1)\Delta(\omega_2)},
\label{eq:chi2_loc}
\end{equation}
with $\Delta(\omega)=\omega_T^2-\omega(\omega+i\gamma_0)$; here $\chi^{(2)}_0$, $\chi^{(2)}_{A,B}$, and $\chi^{(2)}_{C,D,E}$ denote respectively the local, first-order nonlocal ($A,B$), and second-order nonlocal ($C,D,E$) contributions, scaling as $1$, $k$, and $k^2$. As summarized in Table~\ref{tab:dft_summary}, this dimensional estimate reproduces the DFPT values for 3C-SiC, validating it as an inexpensive tool for the broader survey in Fig.~\ref{fig1} (material-specific parameters are given in the Supplementary Information).

We now turn to centrosymmetric crystals, where the nonlocal contribution is the only bulk second-order response. This is a qualitatively new effect: a bulk response activated solely by PhP confinement in materials where inversion symmetry forbids any local $\chi^{(2)}$, growing linearly with $k$. Dimensional analysis,
\begin{equation}
|\chi^{(2)}_0| \sim \frac{\left[\varepsilon_\infty(\omega_L^2-\omega_T^2)\right]^{3/2}}{\omega_T^2\sqrt{\varepsilon_0\rho}},
\label{eq:chi2_ref}
\end{equation}
predicts a reference local scale of $\sim 4$--$5~\mathrm{pm/V}$ for the rocksalt and fluorite crystals considered here. Multiplication by the expected nonlocal factor $ka$ then yields off-resonant nonlocal susceptibilities of $\sim 0.1$ to $0.3~\mathrm{pm/V}$, in good agreement with the DFPT results reported in Table~\ref{tab:dft_summary}.

However, the nonlocal response can be enhanced dramatically near phonon resonances. When the fundamental and second harmonic simultaneously approach phonon poles---the doubly resonant condition $\omega\approx\omega_T$ and $2\omega\approx\omega_L$---the damping-limited phonon denominators produce an enhancement of order $(\omega_T/\gamma_0)^2\sim10^4$, activated only under strong PhP confinement. The resulting upper bound is
\begin{equation}
|\chi^{(2)}_{A,B}|^\mathrm{res}\sim
\frac{ka\left[\varepsilon_\infty(\omega_L^2-\omega_T^2)\right]^{3/2}}
{\omega_T^2\sqrt{\varepsilon_0\rho}}
\left(\frac{\omega_T}{\gamma_0}\right)^2.
\label{eq:chi2_res}
\end{equation}
The degree of enhancement depends on how close a material's phonon spectrum lies to $\omega_L=2\omega_T$. For 3C-SiC, $\omega_L/\omega_T\approx1.21$ places $2\omega_T$ outside the RB, so only a singly resonant enhancement is accessible, giving $|\chi^{(2)}_{A,B}|^\mathrm{res}\simeq560\ \mathrm{pm/V}$ (DA), consistent with DFPT (Table~\ref{tab:dft_summary}). Among all materials surveyed, LiF is the most promising: its ratio $\omega_L/\omega_T\simeq2.15$ is closest to the ideal condition, and its low phonon damping yields resonant values (DA and DFPT, Table~\ref{tab:dft_summary}) that exceed the local $\chi^{(2)}$ of canonical nonlinear crystals such as LiNbO$_3$ ($\sim40\ \mathrm{pm/V}$) or GaAs ($\sim200\ \mathrm{pm/V}$). CaF$_2$ and BaF$_2$ lie closer still to $\omega_L=2\omega_T$ than MgO, but their smaller LO-TO splittings and larger damping yield comparatively modest resonant bounds ($\sim2\times10^3$ and $\sim5\times10^2\ \mathrm{pm/V}$, Fig.~\ref{fig1}). In the limit $\omega_L\to2\omega_T$, an additional enhancement factor $\omega_L/\gamma_0\sim10^2$ could be gained.

\begin{table}[h]
\centering
\caption{\textbf{Comparison of DFPT and dimensional analysis (DA) estimates of the local and nonlocal $\chi^{(2)}$ in centrosymmetric and non-centrosymmetric polar crystals.} Values at $k=150k_0$; DFPT estimates use $\gamma_0=4\ \mathrm{cm}^{-1}$. Second-order nonlocal terms $\chi^{(2)}_{CDE}$, relevant only off-resonance in non-centrosymmetric 3C-SiC, are estimated (DA) at $6$ to $7\times10^{-3}$ (see Supplementary Information).}
\label{tab:dft_summary}
\begin{tabular}{lccccc}
\toprule
& \multicolumn{1}{c}{$|\chi^{(2)}_0|$ (pm/V)}
& \multicolumn{2}{c}{$|\chi^{(2)}_{A,B}|^\mathrm{off\text{-}res}$ (pm/V)}
& \multicolumn{2}{c}{$|\chi^{(2)}_{A,B}|^\mathrm{res}$ (pm/V)} \\
\cmidrule(lr){2-2}\cmidrule(lr){3-4}\cmidrule(lr){5-6}
Material & DA / DFPT & DA & DFPT & DA & DFPT \\
\midrule
3C-SiC & 80 / 83.4 & 0.35 & 9.5$\times10^{-3}$ to 0.35 & 560 & 12.6 to 828 \\
LiF    & Forbidden & 0.3 & 0.060 to 1.48 & 3000 & 115 to 2827 \\
MgO    & Forbidden & 0.1 & $2.9\times10^{-4}$ to 0.28 & 1000 & 0.90 to 863 \\
\bottomrule
\end{tabular}
\end{table}

\section*{Prospects for nonlocal-nonlinear PhPs}
Several directions follow from this work. The most direct experimental test is near-field microscopy or SHG spectroscopy in centrosymmetric crystals, where the linear dependence of the nonlocal response on $k$ provides a clear signature distinguishing it from conventional surface SHG, which is essentially $k$-independent. SFG microscopy may offer an alternative route to probing these nonlocal nonlinearities, given its high spatial resolution and efficient visible-wavelength detection. In non-centrosymmetric crystals, selected polarization or azimuthal configurations can suppress the bulk local $\chi^{(2)}$, allowing the nonlocal contribution to dominate or even make the local response can interfere with the nonlocal term, providing a heterodyne-like amplification that further enhances its detection.

Extending the framework to fully anisotropic crystals would enable studies of hyperbolic vdW materials such as hBN and $\alpha$-MoO$_3$, where directional PhP confinement may further enhance the nonlocal $\chi^{(2)}$ tensor, while twisted moiré polaritonic systems — whose extreme confinement is expected to strengthen these effects — open a further opportunity for the framework developed here \cite{Shi_25_twodimensional,Yao21_enhanced}.

\backmatter

\section*{Methods}

\subsection*{Density-functional perturbation theory calculations}
\label{sec:dft_methods}

We performed density-functional perturbation theory (DFPT) calculations for LiF, MgO, and 3C-SiC using the PBEsol exchange-correlation functional \cite{perdew2008pbesol}. All
calculations used Quantum ESPRESSO (QE) and the D3Q software package
\cite{Giannozzi_2009,Giannozzi_2017,Paulatto_2013}, together with PseudoDojo ONCV norm-conserving pseudopotentials \cite{VANSETTEN201839,Hamann_2013}.
We used a plane-wave kinetic-energy cutoff of 110~Ry, a charge-density cutoff of 880~Ry, and \(16\times16\times16\) \(k\)-point meshes. Harmonic phonon
dispersions, Born effective charges \(Z^*\), high-frequency dielectric
constants \(\varepsilon_\infty\), and nonlinear coefficients \(\beta_T,\beta_L\) were obtained from standard DFPT linear
response. The resulting harmonic quantities are reported in the SI and agree
well with prior PBEsol studies
\cite{abouhaibeh2026lif_fluorides,zhang2023mgo_pbesol,petretto2018ht_dfpt}.

The leading nonlocal tensor was extracted from reciprocal-space D3Q
calculations of the third-order dynamical matrix
\(D^3(\mathbf q,0,-\mathbf q)\) at small momentum-conserving triplets. The
local cubic term was extracted from \(D^3(0,0,0)\). For zincblende SiC this
gives one independent local component, \(g^{(0)}_{xyz}\), while for rocksalt
LiF and MgO the local cubic term vanishes by inversion symmetry. The
mode-weighted optical-coordinate tensor \(A^{\rm opt}\) was obtained by
contracting the atom-resolved third-order response with the optical-mode
weights. Since the D3Q response is used directly at finite \(q\), the extracted
\(A^{\rm opt}\) coefficients should be interpreted as effective
mode-projected finite-\(q\) coefficients in the present computational
convention. In particular, no explicit separation of third-order short-range
anharmonic and long-range polar electrostatic contributions was applied.
Further details of the extraction protocol are given in the SI.

The DFPT-derived local second-order susceptibility of SiC was calculated from
Eq.~\eqref{eq:chi2_loc} using the ab-initio values of \(\mu=eZ^*/V_{\rm cell}\), \(\rho_R\), \(\omega_T\), and \(g^{(0)}_{\alpha\beta\gamma}\) (continuum limit), while the damping is fixed phenomenologically using \(\gamma_0=4~\mathrm{cm}^{-1}\). As for the dimensional analysis estimates, this value was evaluated away from the phonon pole. The finite-$k$ nonlocal susceptibility was estimated from the corresponding projected force coefficient $\Lambda_{pqr}(\hat{q})$. We emphasize that $\hat{q}$ is the propagation direction of the PhP wavevector used in the D3Q triplet, not an optical field-polarization direction, and that $p,q,r \in \{L, T_1, T_2\}$ label phonon-polarization channels relative to $\hat{q}$---distinct from the Cartesian crystal-frame indices $a,b,c$ used for the conventional optical tensor $\chi^{(2)}_{abc}$ (Eq.~4). We adopt the $p,q,r$ notation here specifically to avoid this collision. Recovering the conventional $\chi^{(2)}_{abc}$ lobes/nodes structure from $\Lambda_{pqr}(\hat{q})$ requires an additional contraction of the resulting susceptibility tensor with the actual input and output optical field polarizations; the present $\Lambda$-plot reports the finite-$k$ lattice coefficient one step prior to that optical-geometry projection, and we leave this contraction to future
work.
\begin{equation}
    \left[\chi^{(2)}_{pqr}\right]_{A,B}(2\omega;\omega,\omega)
    \simeq
    \frac{\mu^{3} k\, \Lambda_{pqr}(\hat{q})}
         {\varepsilon_0\, \rho_R^{3}\, \Delta(2\omega)\, [\Delta(\omega)]^{2}},
\end{equation}
where $\Delta(\omega) = \omega_T^2 - \omega(\omega + i\gamma_0)$
(consistent with Eq.~4). The off-resonant values were evaluated for \(k=150k_0\) using \(\Delta(\omega)\approx \Delta(2\omega)\approx\omega_T^2\),
whereas the TO-resonant upper estimates were obtained by setting
\(\omega=\omega_T\). We report selected channel contributions in the
SI, which illustrate up to an order-of-magnitude spread in the susceptibility
depending on the propagation and polarization channel. This spread reflects the physical anisotropy of the projected
rank-four nonlinear tensor as different propagation and polarization combinations
sample different tensor components.

\subsection*{Linear optical response}

The mechanical potential $U[\{X_\alpha\}] = \sum_i U_i[\{X_\alpha\}]$ used in the main text is a functional of all three displacement components, expanded as an infinite series in powers of $X_\alpha$; for compactness we write $U$ in place of $U[\{X_\alpha\}]$. In this work $U$ is taken to depend only on the displacement field, isolating the role of phonon-gradient nonlinearities, while the electromagnetic field couples locally to the lattice through the standard dipolar interaction. More general formulations could incorporate field-gradient nonlinearities directly, via $U[\{X_\alpha\},\mathbf{E}]$; in the complementary limit where the lattice response is instead treated as local and the electromagnetic field carries the nonlocality, such a formulation recovers the familiar multipolar contributions to nonlinear optical response. Truncating $U$ at harmonic order, $U\approx U_2$, gives the linear regime, determined by the harmonic potential
\begin{equation}
    U_2 = \frac{\rho}{2}\int\frac{d^3k}{(2\pi)^3}\,
    \tilde{X}_\alpha(-\mathbf{k})\,\mathcal{D}^{\mathrm{SR}}_{\alpha\beta}(\mathbf{k})\,\tilde{X}_\beta(\mathbf{k}),
    \label{eq:Uharm_general}
\end{equation}

Equation \eqref{eq:Uharm_general} is the wavevector-space generalization of Hooke's law~\cite{Born54_Dynamical}. Each Fourier mode $\tilde{X}_\alpha(\mathbf{k})$ experiences a restoring force proportional to its own amplitude, with stiffness tensor $\mathcal{D}^{\mathrm{SR}}_{\alpha\beta}(\mathbf{k})$. The full dynamical matrix splits as $\mathcal{D}_{\alpha\beta}(\mathbf{k}) = \mathcal{D}^{\mathrm{SR}}_{\alpha\beta}(\mathbf{k}) + \mathcal{D}^{\mathrm{LR}}_{\alpha\beta}(\mathbf{k})$, where the short-range part $\mathcal{D}^{\mathrm{SR}}_{\alpha\beta}$ encodes finite-range interatomic forces decaying over a few bond lengths, while the long-range part $\mathcal{D}^{\mathrm{LR}}_{\alpha\beta}(\mathbf{k})$ is the Coulomb contribution. We exclude $\mathcal{D}^{\mathrm{LR}}_{\alpha\beta}$ from $U_2$ and instead absorb it into the total electric field $E_\alpha$, as discussed below.

Expanding $\mathcal{D}^{\mathrm{SR}}_{\alpha\beta}(\mathbf{k})$ in powers of $k^2$---odd powers being absent by time-reversal symmetry---and truncating at $\mathcal{O}(k^2)$ as the 
minimal extension capturing spatial dispersion gives
$\mathcal{D}^{\mathrm{SR}}_{\alpha\beta}(\mathbf{k}) \approx
\mathcal{D}^{\mathrm{SR}}_{0,\alpha\beta} + \Sigma_{\alpha\beta\gamma\delta}\, k_\gamma k_\delta$,
where $\mathcal{D}^{\mathrm{SR}}_{0,\alpha\beta}$ is the $k=0$ short-range force constant tensor, whose eigenvalues give the zone-center TO frequencies, and $\Sigma_{\alpha \beta \gamma \delta}$ is the fourth-rank tensor of spatial dispersion coefficients, encoding how the restoring forces stiffen with wavevector. Transforming $\mathcal{D}^{\mathrm{SR}}_{\alpha\beta}(\mathbf{k})$ to real space via $k_\gamma k_\delta \tilde{X}_\beta(\mathbf{k}) \leftrightarrow -\partial_\gamma\partial_\delta X_\beta(\mathbf{r})$, the harmonic potential becomes
$U_2 = \frac{\rho}{2}\int d^3r\,[
X_\alpha\,\mathcal{D}^{\mathrm{SR}}_{0,\alpha\beta}\, X_\beta + (\partial_\gamma X_\alpha)\,\Sigma_{\alpha\beta\gamma\delta}\,(\partial_\delta X_\beta) ]$. Varying $U_2$ with respect to $X_\alpha$ and substituting into Eq.~\eqref{euler_lagrange}, then assuming time-harmonic motion $X_\alpha \propto e^{-i\omega t}$ so that $\ddot{X}_\alpha \to -\omega^2 X_\alpha$ and $\dot{X}_\alpha \to -i\omega X_\alpha$, yields the general equation of motion
\begin{equation}
    \bigl[-\omega(\omega+i\gamma_0)\delta_{\alpha\beta}
    + \mathcal{D}^{\mathrm{SR}}_{0,\alpha\beta}
    - \Sigma_{\alpha\beta\gamma\delta}\,\partial_\gamma\partial_\delta\bigr]X_\beta
    = \frac{\mu}{\rho} E_\alpha,
    \label{eq:eom_general}
\end{equation}
where $E_\alpha$ is the total electric field, including the depolarization field generated by the ionic displacement itself. It is through this field that the long-range Coulomb term excluded from $U_2$ re-enters the dynamics: for a longitudinal displacement, the depolarization field contributes precisely $\mathcal{D}^{\mathrm{LR}}_{\alpha\beta}X_\beta$ to the restoring force, and it is this that gives rise to the LO-TO splitting. Equation \eqref{eq:eom_general} is general, applicable to crystals of arbitrary symmetry and, upon inclusion of higher-order terms in $U$, to nonlinear regimes.

For an isotropic crystal $\mathcal{D}^{\mathrm{SR}}_{0, \alpha\beta} = \omega_T^2 \delta_{\alpha \beta}$ and the fourth-rank tensor $\Sigma_{\alpha\beta\gamma\delta}$ reduces to two independent scalar coefficients $\beta_L^2$ and $\beta_T^2$, so that $\mathcal{D}^{\mathrm{SR}}_{\alpha\beta}(\mathbf{k})$ becomes
$\mathcal{D}^{\mathrm{SR}}_{\alpha\beta}(\mathbf{k}) = \omega_T^2\,\delta_{\alpha\beta} + \beta_T^2 k^2\!\left(\delta_{\alpha\beta} -\hat{k}_\alpha \hat{k}_\beta\right) + \beta_L^2 k^2\,\hat{k}_\alpha\hat{k}_\beta$, where $\hat{k}_\alpha\hat{k}_\beta$ projects onto the longitudinal direction and  $\delta_{\alpha\beta}-\hat{k}_\alpha\hat{k}_\beta$ onto the transverse plane. The long-range contribution takes the explicit form $\mathcal{D}^{\mathrm{LR}}_{\alpha\beta}(\mathbf{k}) = 
(\omega_L^2-\omega_T^2)\hat{k}_\alpha\hat{k}_\beta = 
(\omega_L^2-\omega_T^2)k_\alpha k_\beta/k^2$. The $k^2$ in the denominator confirms the non-analyticity at $\mathbf{k}=0$: its value depends on the direction from which $\mathbf{k}\to 
0$, reflecting the fact that only a longitudinal displacement builds up a macroscopic charge density and hence a long-range depolarization field. It is through this field, entering via 
$E_\alpha$, that the LO-TO splitting $\omega_L^2 = \omega_T^2 + 
\mu^2/(\varepsilon_0\varepsilon_\infty\rho)$ is recovered.

Substituting $\mathcal{D}^{\mathrm{LR}}_{\alpha\beta}(\mathbf{k})$ back into $U_2$, the harmonic potential then becomes $U_2 = \frac{1}{2}\rho\int d^3r\,[\omega_T^2 X_\alpha X_\alpha + \beta_L^2(\partial_\alpha X_\alpha)^2 \nonumber \\+\beta_T^2\,\epsilon_{\alpha\beta\gamma}(\partial_\beta X_\gamma)^2]$, where $\epsilon_{\alpha\beta\gamma}$ is the Levi-Civita tensor. Varying with respect to $X_\alpha$ and substituting into Eq.~\eqref{euler_lagrange} yields 
$\bigl[\omega_T^2\delta_{\alpha\beta} -\omega(\omega + i\gamma_0)\delta_{\alpha\beta}+ \beta_L^2 \partial_\alpha \partial_\beta-\beta_T^2\,\epsilon_{\alpha \gamma \delta}\epsilon_{\beta \nu \delta} \partial_\gamma \partial_\nu\bigr]X_\beta = \frac{\mu}{\rho} E_\alpha$. Recognizing $\partial_\alpha \partial_\beta \equiv [\nabla(\nabla\cdot)]_{\alpha\beta}$ and $\epsilon_{\alpha\gamma\delta} \epsilon_{\beta\nu\delta} \partial_\gamma\partial_\nu \equiv -[\nabla\times \nabla \times]_{\alpha\beta}$, this is equivalently written in the compact 
vector form
\begin{equation}
\bigl[\omega_T^2 - \omega(\omega+i\gamma_0)
  + \beta_L^2\nabla(\nabla\cdot)
  - \beta_T^2\nabla\times\nabla\times\bigr]\mathbf{X}
= \frac{\mu}{\rho}\mathbf{E},
\label{eq:eom}
\end{equation}
where $\mathbf{X}$ and $\mathbf{E}$ collect the Cartesian components $X_\alpha$ and $E_\alpha$, 
respectively. Equation~\eqref{eq:eom} recovers the result of Ref.~\cite{Gubbin20_optical}. The 
phonon velocities $\beta_L$ and $\beta_T$ measure how fast longitudinal and transverse lattice 
disturbances propagate through the crystal, and are the phonon analogues of the electron Fermi velocity. This leads to nonlocality and a wavevector-dependent dielectric function $\varepsilon(\omega,\mathbf{k})$ beyond the local $\varepsilon(\omega)$.
The condition $\varepsilon_L(\omega, \mathbf{k})=0$ therefore defines a dispersive LO phonon branch at $\omega$ and (complex) wavevectors $\mathbf{k}$ \cite{Gubbin20_optical}.

The macroscopic polarization is $\mathbf{P} = \mu\mathbf{X} + \varepsilon_0(\varepsilon_\infty-1)\mathbf{E}$, 
where the first term arises from ionic motion and the second from the instantaneous electronic response. Combined with the wave equation $\nabla\times\nabla\times\mathbf{E}-\frac{\omega^2}{c^2}\varepsilon_\infty\mathbf{E}=\mu_0\omega^2\mathbf{P}$, 
Eq.~\eqref{eq:eom} forms a closed coupled system that can be solved numerically for arbitrary geometries. This framework 
is directly analogous to the nonlocal hydrodynamic description of electron gases in 
plasmonics, 
where analogous spatial-gradient terms are known to underpin strong and tunable nonlinear optical responses \cite{de2021free,hu2024low,Rossetti2025_control,Hu25_Modulating,AlvarezPerez2025_ultrahigh}. It is this type of nonlocal-nonlinear sources that we uncover in this work.

\bmhead{Data availability}

The data that support the findings of this study are available from the corresponding authors upon reasonable request.

\bmhead{Code availability}

The code that supports the plots and data analysis of this study is available from the corresponding authors upon reasonable request.
 
\bmhead{Author contributions} 

G.Á.-P. conceived, led and supervised the study, developed the theoretical framework, and wrote the initial draft of the manuscript. L.J. performed the DFPT calculations. S.D.L. contributed to the conception of the study and the development of the theoretical framework. All authors contributed to the interpretation and discussion of the results and participated in the preparation of the manuscript.

\bmhead{Funding}
 
G.Á.-P. would like to thank Cristian Ciracì, Ion Errea and Jaime Ferrer for helpful discussions. G.Á.-P. acknowledges support from the European Union (Marie Skłodowska-Curie Actions, grant agreement No. 101209198). L.J. is supported by a career grant from the Novo Nordisk Foundation (grant no. NNF25OC0105410). The Center for Polariton-driven Light-Matter Interactions (POLIMA) is funded by the Danish National Research Foundation (Project No. DNRF165). S.D.L. acknowledges financial support under the National Recovery and Resilience Plan (NRRP), Mission 4, Component 2, Investment 1.1, Call for tender No. 1409 published on 14/09/2022
by the Italian Ministry of University and Research (MUR), funded by the European Union – NextGenerationEU – Project
Title MINAS - CUP B53D23028420001 - Grant Assignment Decree No. 1380 adopted on 01/09/2023 by the Italian Ministry of University and Research (MUR).

\bmhead{Competing interests}
 
The authors declare no competing interests.

\bibliography{mybib}

\end{document}